\documentclass[twocolumn,trackchanges]{aastex62} 

\submitjournal{ApJ}

\shorttitle{Cosmological tests of gravity with latest observations}
\shortauthors{Li \& Zhao}

\def\be{\begin{equation}}
\def\ee{\end{equation}}
\def\ba{\begin{eqnarray}}
\def\ea{\end{eqnarray}}

\def\ie{{\frenchspacing\it i.e.}}
\def\eg{{\frenchspacing\it e.g.}}

\def\nn{\nonumber}

\def\l{\left}
\def\r{\right}

\def\lnB{{\rm log}_{10}B_0}
\def\hoMpc{h^{-1} {\rm Mpc}}

\begin{document}

\title{Cosmological tests of gravity with latest observations}

\author{Jian Li}
\affiliation{University of Chinese Academy of Sciences, Beijing, 100049, P.R.China}
\affiliation{National Astronomy Observatories, Chinese Academy of Science, Beijing, 100101, P.R.China}

\author[0000-0003-4726-6714]{Gong-Bo Zhao}
\affiliation{National Astronomy Observatories, Chinese Academy of Science, Beijing, 100101, P.R.China}
\affiliation{University of Chinese Academy of Sciences, Beijing, 100049, P.R.China}
\affiliation{Institute of Cosmology and Gravitation, University of Portsmouth, Portsmouth, PO1 3FX, UK}

\begin{abstract}
We perform observational tests of modified gravity on cosmological scales following model-dependent and model-independent approaches using the latest astronomical observations, including measurements of the local Hubble constant, cosmic microwave background, the baryonic acoustic oscillations and redshift space distortions derived from galaxy surveys including the SDSS BOSS and eBOSS, as well as the weak lensing observations performed by the CFHTLenS team. Using all data combined, we find a deviation from the prediction of general relativity in both the effective Newton's constant, $\mu(a,k)$, and in the gravitational slip, $\eta(a,k)$. The deviation is at a $3.1\sigma$ level in the joint $\{\mu(a,k),\eta(a,k)\}$ space using a two-parameter phenomenological model for $\mu$ and $\eta$, and reaches a $3.7\sigma$ level if a general parametrization is used. This signal, which may be subject to unknown observational systematics, or a sign of new physics, is worth further investigating with forthcoming observations.
\end{abstract}

\keywords{Cosmology: modified gravity; dark energy}

\section{Introduction} \label{sec:intro}

The physical law governing the accelerating expansion of the Universe, which was discovered by the redshift-luminosity relation revealed from supernovae observations \citep{Riess, Perlmutter}, remains unveiled. In principle, the cosmic acceleration may suggest that approximately two-thirds of the total energy budget of the current Universe is provided by an unknown energy component with a negative pressure, dubbed dark energy (DE) \citep{DEreview1,DEreview2}, or that we need a better understanding of the law of gravity. 

The cosmological constant (CC) or the vacuum energy $\Lambda$, introduced by Einstein a century ago to prevent the Universe from collapsing, has ironically become one of the most popular candidates of DE to give rise to the cosmic acceleration. Although the $\Lambda$-cold dark matter ($\Lambda$CDM) model can fit observations reasonably well, it suffers from severe theoretical issues \citep{CCissue}. Dynamical dark energy models \citep{DEreview2} can alleviate the cosmological constant problem to some extent, and phenomenological approaches in light of observations have been developing actively (see \citealt{ZhaoDE17} for an example).

On the other hand, general relativity (GR) is the most successful theory of gravity on scales from laboratory to the solar system. However, the validity of GR on cosmological scales is postulated, which is subject to scrutiny in theory, and to tests in observations. In fact, the expansion of the Universe can accelerate without the existence of dark energy, if the left-hand side of the Einstein equation gets modified. This essentially alters the response of the spacetime curvature to the energy-momentum distribution, and it is dubbed the modified gravity (MG) scenario (see \citealt{MGreview1,MGreview2,MGreview3,MGreview4} for reviews on MG). 

Both dark energy and modified gravity can yield the same expansion history of the Universe after the required tuning of parameters, however, these two scenarios predict different growth of the cosmic structures. In other words, DE and MG can be degenerate at the background level, but this `dark degeneracy' can be broken at the perturbation level \citep{DarkD}.

Given our ignorance of the nature of dark energy and gravity, every possibility is worth exploring. In this regard, a combination of multiple cosmic probes, which is able to determine the cosmic expansion and structure growth history simultaneously, plays a key role for DE and MG studies.

In this work, we focus on observational tests of modified gravity scenarios on linear scales, on which the linear perturbation theory is valid. On these scales, MG can change the effective Newton's constant and/or the geodesics of photons \citep{MGreview1}, which leaves imprints on various kinds of cosmological observations, including the cosmic microwave background (CMB) and large-scale structure (LSS) of the Universe. In particular, redshift space distortions (RSD) \citep{RSD1,RSD2} derived from the galaxy clustering of LSS spectroscopic surveys probe the change in the effective Newton's constant. Weak lensing (WL) measured from the imaging LSS surveys constrains the deviation of photon's trajectory from the geodesic in a flat space, making RSD and WL highly complementary to each other for gravity tests \citep{MGPVWL,MGPLC,MGCFHTLENS,MGobsZhao1,MGobsZhao2}.

In this analysis, we use the latest observations of CMB and LSS, combined with background cosmology probes, to derive constraints on modified gravity scenarios in a phenomenological way. Those background probes include the local measurement of the Hubble constant ($H_0$), the Hubble rate measurements using passive galaxies (OHD), and baryonic acoustic oscillations (BAO) \citep{BAO1,BAO2}.

The paper is structured as follows.  Section \ref{sec:method} describes the methodology used for this analysis, including the observational datasets, the rationale and framework of parametrizations of modified gravity, and details of the parameter estimation procedure. Our main results are presented in Section \ref{sec:results}, before conclusion and discussions in Section \ref{sec:conclusion}.

\section{Methodology}
\label{sec:method}

In this section, we present the methodology used for this analysis, including the general framework in which we parametrize the effect of modified gravity, datasets used, and details for parameter estimation.

\subsection{General framework of parametrizing modified gravity} 
\label{sec:para}

In this section, we discuss how we parametrize the Universe in gravity models beyond GR. As we aim to use the growth of cosmic structure to break the dark degeneracy between MG and DE, in this work we assume a $\Lambda$CDM background cosmology, and parametrize the modification of gravity at the linear perturbation level.

In a flat Friedmann-Robertson-Walker (FRW) Universe, the metric in 
the conformal-Newtonian gauge reads, 
\be\label{metric}
ds^2=-a^2(\tau)\l[\l(1+2\Psi\r)d\tau^2-\l(1-2\Phi\r) d\vec{x}^2\r] \ ,
\ee
where $\Phi$ and $\Psi$ are functions depending on time (redshift $z$) and scale (wavenumber $k$). The energy-momentum conservation yields,
\ba
\label{eq:continuity}
\delta'+{k\over aH}v-3\Phi'&=&0, \nonumber \\
v'+v-{k\over aH}\Psi&=&0. \ 
\ea where $\delta$ refers to the density contrast, $v$ represents the irrotational component of peculiar velocity, $a$ and $H$ are the scale factor and the Hubble rate respectively, and the prime denotes derivatives with respect to ${\rm ln} \ a$.

In order to solve for $\{\delta, v, \Psi, \Phi\}$, two additional equations are required to close the system, and this is where a theory of gravity is required. Generically, the required equations are as follows \citep{MGCAMB2,MGParaLP} \footnote{Alternative frameworks for parametrizing modified gravity have been proposed, \eg, \cite{MGParaOx} and the effective field theory approach developed in \citet{EFTCAMB1,EFTCAMB2}.},
\ba
\label{eq:Poisson}
k^2\Psi&=&-4\pi G \rho a^2 \mu(a,k) \Delta, \\
\label{eq:gslip}\frac{\Phi}{\Psi}&=&\eta(a,k).
\ea
where Eqs. (\ref{eq:Poisson}) and (\ref{eq:gslip}) are called the modified Poisson equation and the gravitational slip equation respectively. $\Delta$, which is defined as $\delta + 3{aH}v/k$, denotes the gauge-invariant, comoving density contrast.

GR predicts that $\mu(a,k)=\eta(a,k)=1$, and any deviation of these functions from unity may be regarded as a smoking gun for modified gravity. Note that the $\mu(a,k)$ function can only be tested on sub-horizon scales, as it becomes irrelevant on super-horizon scales, on which only $\eta(a,k)$ can be tested observationally. On sub-horizon scales, both $\mu(a,k)$ and $\eta(a,k)$ have observational effects to be tested.

 As Big Bang Nucleosynthesis (BBN) and CMB have been well explained with theories based on GR, we assume GR at high redshifts by setting $\mu(a,k)=\eta(a,k)=1$ at $z>50$, and test the deviation of $\mu$ and $\eta$ from unity at lower redshifts.

Before introducing specific MG models to be tested, we parameterize our Universe with the following set of cosmological parameters, 
\be
\label{eq:paratriz} {\bf P} \equiv (\Omega_{\rm b}h^{2}, \Omega_{\rm cdm}h^{2}, \Theta_{s}, \tau, n_s, A_s, \mathcal{N}, \mathcal{X}) \ee where $\Omega_{\rm b}h^{2}$ and $\Omega_{\rm cdm}h^{2}$ denote the physical baryon and cold dark matter energy density, respectively; $\Theta_{s}$ is the ratio ($\times100$) between the sound horizon and the angular diameter distance at last scattering surface; $\tau$ is the re-ionisation optical depth; and $n_s$ and $A_s$ denote the primordial power spectrum index and the amplitude of primordial power spectrum, respectively. In addition, $\mathcal{N}$ is used to denote several nuisance parameters that will be marginalized over when performing the likelihood analysis, and $\mathcal{X}$ denotes parameters to parametrize the $\mu(a,k)$ and $\eta(a,k)$ functions. As we only test gravity at the perturbation level, we assume a flat $\Lambda$CDM background cosmology.

\subsection{Datasets}

\begin{table*}[htp]
\begin{center}
\begin{tabular}{c|c|c}

\hline\hline
Measurements & Meaning & References \\
\hline
PLC & CMB provided by the Planck collaboration & \cite{Planck15} \\
6dF & BAO using the 6dFGS survey & \cite{6dFBAO} \\ 
MGS &  BAO from the SDSS MGS sample & \cite{MGS} \\
Ly$\alpha$FB & BAO from the SDSS DR11 Ly$\alpha$-forest sample & \cite{LyaFB} \\
Alam & Consensus BAO + RSD using the BOSS DR12 combined sample & \cite{Alam17} \\
Wang & Tomographic BAO + RSD using the BOSS DR12 combined sample & \cite{Wang17RSD} \\
eBOSS & Tomographic BAO + RSD using the eBOSS DR14 quasar sample & \cite{Zhao18} \\
SNe & Luminosity from the JLA supernovae sample & \cite{JLA} \\ 
$H_0$ & Recent local $H_0$ & \cite{R16} \\
WL & Weak lensing shear using the CFHTLenS sample & \cite{WL} \\
$P(k)$ & Power spectrum from WiggleZ & \cite{WiggleZ}\\
OHD & $H(z)$ using the ages of passive galaxies & \cite{OHD} \\

\hline
BAORSD & 6dF+MGS+Ly$\alpha$FB+Wang &    \\
BSH & BAORSD+SNe+$H_0$+OHD&   \\
ALL17 & PLC+BSH+WL+$P(k)$ &  \\
ALL18 & ALL17+eBOSS & \\
\hline\hline
\end{tabular}
\end{center}
\caption{List of datasets used for this analysis, with acronyms, meanings and references.}
\label{tab:data}
\end{table*}%

The observational datasets used for this analysis include the cosmic microwave background (CMB), supernovae (SNe), BAO \& RSD, weak lensing (WL), galaxy power spectrum, and observational $H(z)$ data (OHD). 

For CMB, we use the angular power spectra from the temperature and polarisation maps provided by the Planck mission \citep{Planck15}. The BAO-alone measurements we use include the isotropic BAO distance estimates using the 6dFGS \citep{6dFBAO} and the Main Galaxy Sample (MGS) of Sloan Digital Sky Survey (SDSS) Data Release (DR) 7 \citep{MGS}, and the anisotropic BAO measurement using the Lyman-$\alpha$ forest in BOSS DR11 \citep{LyaFB}. For joint BAO and RSD, we use three recent measurements including，
\begin{itemize}
\item The consensus measurement at three effective redshifts of $z=\{0.38, 0.51, 0.61\}$ using the BOSS DR12 combined sample \citep{Alam17};
\item The tomographic BAO and RSD measurement at nine effective redshifts in the range of $z\in[0.2,0.75]$ derived from the same DR12 sample \citep{Wang17RSD}\footnote{Note that as (I) and (II) are derived from the same galaxy sample, we use them separately in our analysis.};
\item The tomographic BAO and RSD measurement at four effective redshifts using the eBOSS \citep{eBOSS1,eBOSS2} DR14 quasar sample based on the optimal redshift weighting method \citep{Zhao18}.
\end{itemize}

Other observational data used for this analysis include the luminosity measurements from the joint light-curve analysis (JLA) SNe sample \citep{JLA}, the recent local $H_0$ measurement \citep{R16}, the weak lensing shear measurement from the CFHTLenS survey \citep{WL}, the galaxy power spectrum measurement from the WiggleZ redshift survey \citep{WiggleZ}, and a compilation of $H(z)$ measurements using the ages of passive galaxies \citep{OHD}.

To be explicit, we make a list of these datasets with acronyms, meanings and references in Table \ref{tab:data}, and will use the acronyms shown in this table for later reference when presenting our results.

\subsection{Parameter estimation}

Given a set of parameters in Eq. (\ref{eq:paratriz}), and the functional forms relating parameters $\mathcal{X}$ to the $\mu(a,k)$ and $\eta(a,k)$ functions, which will be introduced in Section \ref{sec:results}, we use {\tt MGCAMB} \citep{MGCAMB1,MGCAMB2} \footnote{Available at \url{http://aliojjati.github.io/MGCAMB/}}, a variant of {\tt CAMB} \citep{CAMB} \footnote{Available at \url{https://camb.info/}} working for modified gravity theories, to compute the observables, and use a modified version of {\tt CosmoMC} \citep{cosmomc} \footnote{Available at \url{https://cosmologist.info/cosmomc/}} to sample the parameter space using the Monte Carlo Markov Chain (MCMC) method.

\section{Results} 
\label{sec:results}

We present our results in this section. To be clear, we present the `scale-independent' and `scale-dependent' cases separately, in which the $\mu$ and $\eta$ functions depend on redshift $z$ only, and on both redshift $z$ and wavenumber $k$. For each case, we explicitly show the parametrization for the $\mu$ and $\eta$ functions, before presenting the observational constraints. We also perform a principal component analysis (PCA) in both cases, to help interpret the result.

\subsection{The scale-independent case}

In this subsection, we consider MG scenarios in which the growth is scale-independent, \ie, $\mu$ and $\eta$ are only functions of time, namely, 
\ba \mu = \mu(a); \ \eta = \eta(a). \ea
We then parametrize the $\mu(a)$ and $\eta(a)$ functions using the gravitational growth index, power-law functions, and a more general parametrization based on piecewise constant bins in redshift.  

\subsubsection{A single parameter extension: the gravitational growth index}
\label{sec:gammaL}

As one of the minimal extensions to GR, the gravitational growth index $\gamma_L$ \citep{gammaL} has been widely used to search for signs of modified gravity phenomenologically (see \citealt{EMM18,Wang17RSD,HGM18,Zhao18} for recent observational tests of gravity using $\gamma_L$). The gravitational growth index is defined as,
\ba
\label{eq:gammaL}
f(a)&\equiv&\frac{{\rm dlog}\delta}{{\rm dlog}a}=\Omega_{\rm M}^{\gamma_L} (a),
\ea where $f(a)$ denotes the logarithmic growth rate as a function of scale factor $a$, $\delta$ is the matter over-density, and $\Omega_{\rm M}(a)$ is the fractional energy density of matter at scale factor $a$.

\begin{figure}[htp]
\includegraphics[scale=0.8]{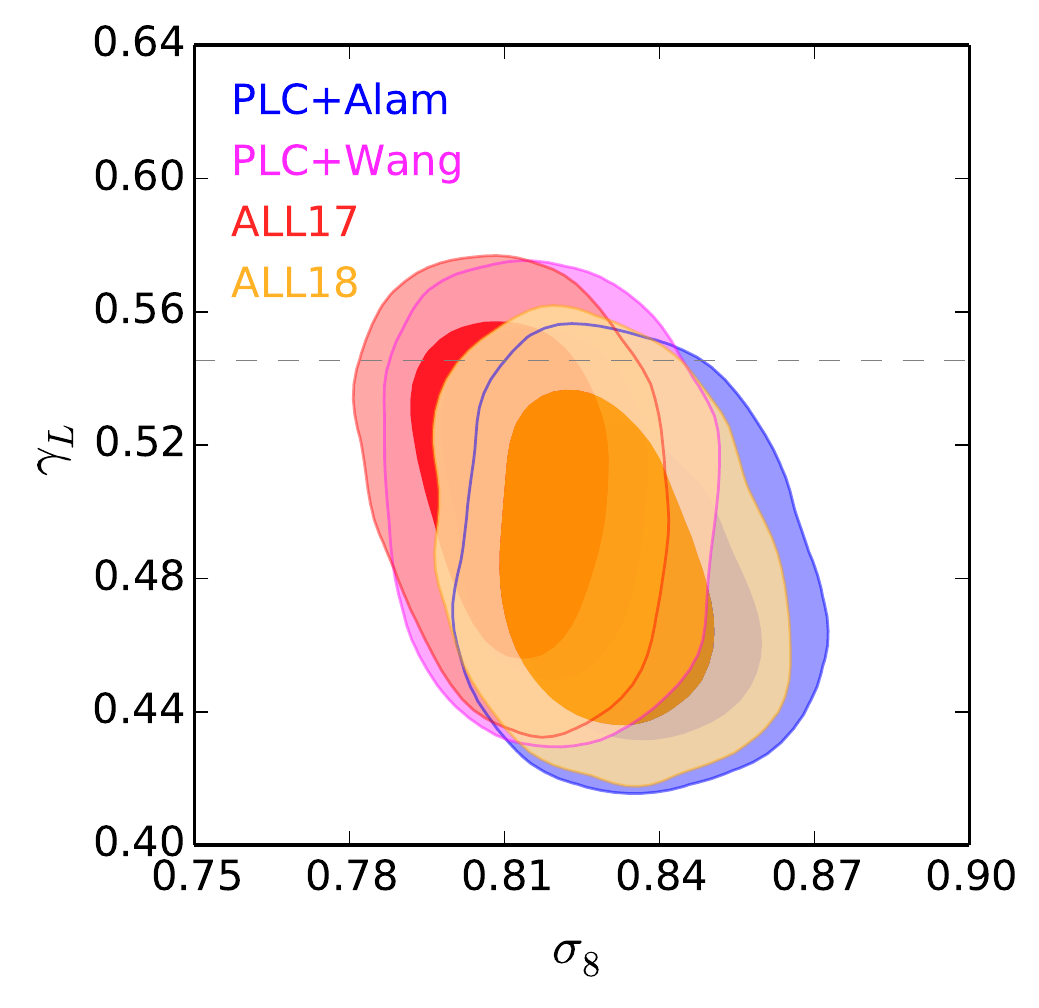}
  \caption{68 and 95\% CL (Confidence Level) contour plots for $\gamma_L$ and $\sigma_8$ derived from four data combinations as illustrated in the legend. The horizontal dashed line shows the GR value of $\gamma_L=0.545$.}
\label{fig:gammaL}	
\end{figure}

\begin{table}[htp]
\begin{centering}
\begin{tabular}{c|c|c}
\hline\hline
    &{$\gamma_L$} & {$\sigma_8$} \\  \hline   
{PLC+Alam} & {$0.478\pm0.029$}& {$0.835\pm0.015$}   \\ \hline 
{PLC+Wang} & {$0.506\pm0.032$}&{$0.818\pm0.013$}   \\ \hline
{ALL17} & {$0.509\pm0.031$} & {$0.812\pm0.013$}   \\ \hline
{ALL18} & {$0.485\pm0.031$} & {$0.828\pm0.014$}   \\ \hline\hline

\end{tabular}
\caption{The mean and 68\% CL uncertainty for parameters $\gamma_L$ and $\sigma_8$ derived from four data combinations.}
\label{tab:gammaL}
\end{centering}
\end{table}

In this framework \citep{MGParaLP} \footnote{Here we omit the variable $a$ for $\Omega_{\rm M}$ for brevity. Also note that this formula is only valid for a constant $\gamma_L$ in a $\Lambda$CDM background. For general cases, \eg, a time-dependent $\gamma_L$ in a general cosmology, see \cite{MGParaLP}.}, 
\ba 
\mu(a)&=& \frac{2}{3}\Omega_{\rm M}^{\gamma_L-1} \left[\Omega_{\rm M}^{\gamma_L}+2-3\gamma_L+3\left(\gamma_L-\frac{1}{2}\right)\Omega_{\rm M}\right]\nonumber \\
\eta(a)&=&1.
\ea

The joint constraints on $\gamma_L$ and $\sigma_8$ (with all other parameters marginalized over) are shown in Table \ref{tab:gammaL} and Fig. \ref{fig:gammaL} for four data combinations. As shown, the GR prediction of $\gamma_L=0.545$ is generally consistent with the observations within the 95\% CL.

\begin{figure*}[htp]
\includegraphics[width=0.99\textwidth]{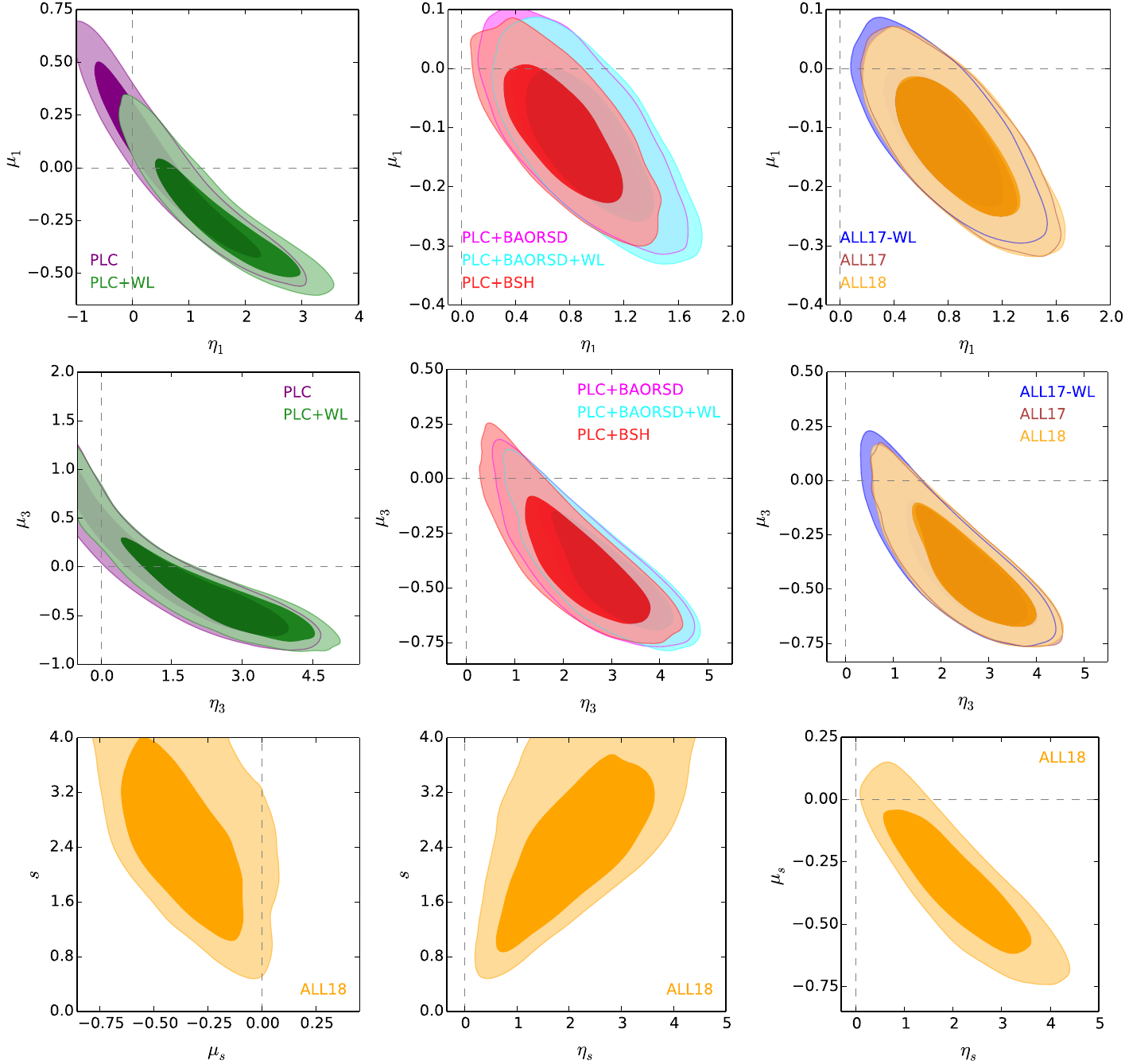}
  \caption{Upper and middle panels: 68 and 95\% contour plots for $\mu_s$ and $\eta_s$, where the upper panels for $s=1$ and middle panels for $s=3$. Contours for different data combinations are shown in separate panels to avoid confusion. Lower panels: 68 and 95\% CL contour plots for $\mu_s$ and $s$ (left) and for $\eta_s$ and $s$ (right) derived from ALL18. In all panels, horizontal and vertical dashed lines denote $\mu_s=0$ and $\eta_s=0$ respectively, and the intersections of the dashed lines denote the GR model.}
\label{fig:powerlaw}	
\end{figure*}

\begin{figure}
\includegraphics[width=0.45\textwidth]{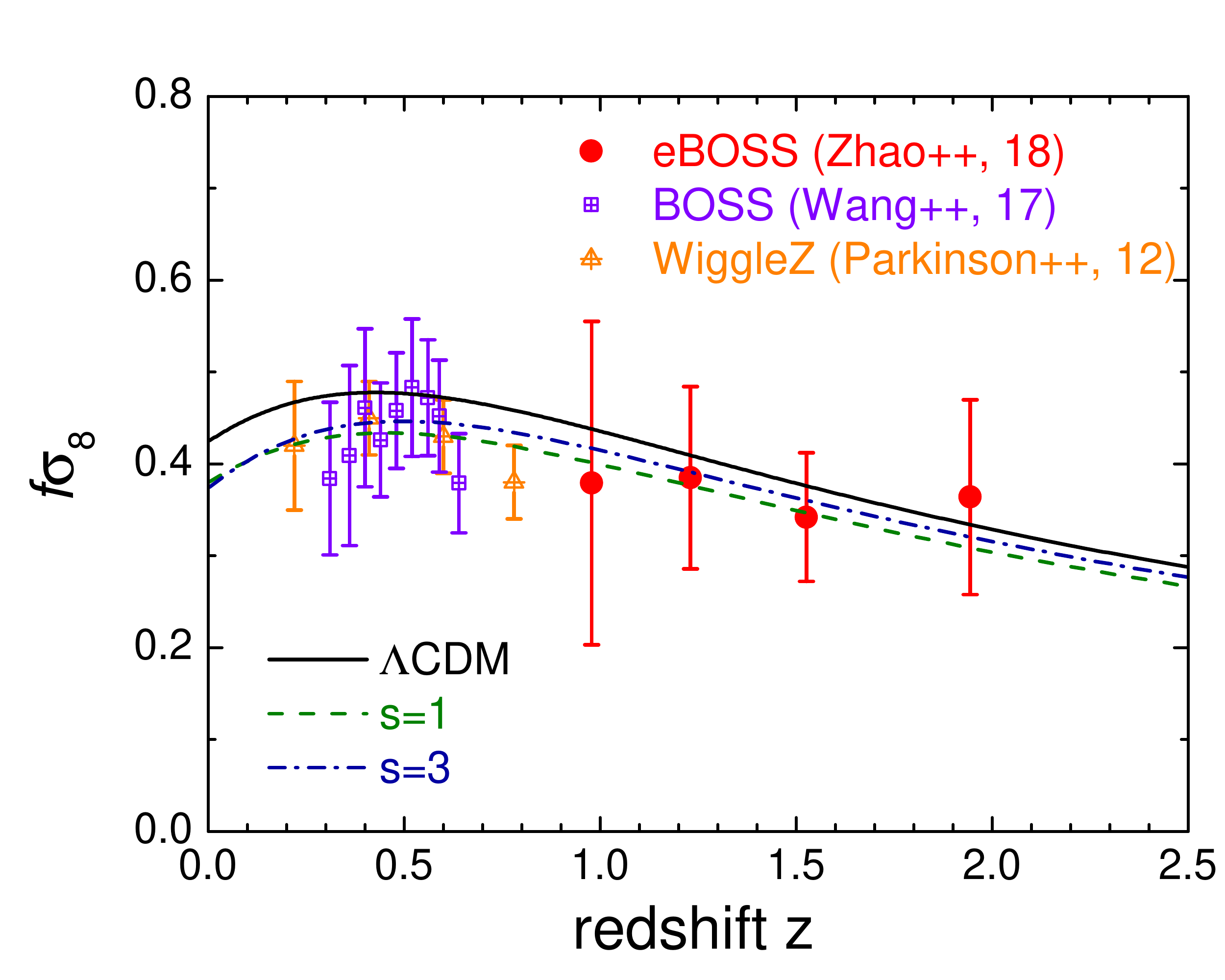}
  \caption{The best-fit $f\sigma_8$ of three gravity models including $\Lambda$CDM (black solid), power-law with $s=1$ (green dashed) and $s=3$ (blue dash-dotted), over-plotted with the observational data as illustrated in the legend.}
\label{fig:fs8}	
\end{figure}

\begin{table*}[htp]
\begin{centering}
\begin{tabular}{c|c|c|c|c}
 \hline\hline
    & \multicolumn{2}{c|}{$s=1$} &
\multicolumn{2}{c}{$s=3$} \\  
  \cline{2-5}
    & $\mu_1$ & $\eta_1$ & $\mu_3$ & $ \eta_3$ \\  \hline
    \multicolumn{1}{c|} {PLC} &
$0.008\pm0.280$ & $0.642\pm0.903$ & $0.008\pm0.538$ &$1.465\pm1.477$\\  \hline
    \multicolumn{1}{c|} {PLC+WL} &
$-0.238\pm0.194$ & $1.546\pm0.795$ & $-0.197\pm0.471$ & $2.218\pm1.428$\\  \hline
    \multicolumn{1}{c|} {PLC+BAORSD} &
$-0.111\pm0.083$ & $0.855\pm0.303$ & $-0.404\pm0.189$ & $2.637\pm0.857$\\  \hline
    \multicolumn{1}{c|} {PLC+BAORSD+WL} &
$-0.128\pm0.082$ & $0.955\pm0.301$ & $-0.429\pm0.175$ & $2.858\pm0.827$\\  \hline
    \multicolumn{1}{c|} {PLC+BSH} &
$-0.110\pm0.074$ & $0.757\pm0.277$ & $-0.371\pm0.195$ & $2.298\pm0.863$\\  \hline
    \multicolumn{1}{c|} {ALL17} &
$-0.131\pm0.075$ & $0.863\pm0.286$ & $-0.405\pm0.184$ & $2.555\pm0.835$\\  \hline
    \multicolumn{1}{c|} {ALL17-WL} &
$-0.114\pm0.074$ & $0.773\pm0.280$ & $-0.376\pm0.197$ & $2.327\pm0.847$\\  \hline
    \multicolumn{1}{c|} {ALL18} &
$-0.132\pm0.075$ & $0.873\pm0.289$ & $-0.398\pm0.184$ & $2.516\pm0.832$ \\  \hline\hline
\end{tabular}
\caption{Mean and 68\% CL uncertainties of the power-law model parameters derived from different data combinations.}
\label{tab:powerlaw}
\end{centering}
\end{table*}

\subsubsection{A three-parameter extension: the power-law parametrization}
\label{sec:powerlaw}

A more general parametrization for $\mu(a)$ and $\eta(a)$ is to use power-law functions \citep{MGobsZhao1},
\ba\label{eq:powerlaw} \mu(a) &=& 1+\mu_s a^s, \nonumber \\
 \eta(a) &=& 1+\eta_s a^s.\ea We consider three cases where $s$ is fixed to $1$ (the linear model), $3$ (the cubic model), or treated as a free parameter to be marginalized over.
 
We constrain the power-law model parameters using various data combinations, and show the result in Table \ref{tab:powerlaw} and Fig. \ref{fig:powerlaw}. As shown, the results for the cases of $s=1$ and $s=3$ are qualitatively similar, so we present both cases together. With PLC alone, GR is excluded at 95\% CL, and adding WL drags the contours towards a direction in which a large positive $\eta_s$ and negative $\mu_s$ are favored (note that $\mu_s=\eta_s=0$ for GR in our notation), which further excludes the GR model. With BAORSD, WL, SNe, and $H_0$ combined with PLC, the contours for both $s=1$ and $s=3$ cases shrink significantly, and GR is excluded beyond the 95\% CL level. Finally, combining all data, denoted as ALL18, yields the tightest constraint, which excludes the GR model at $2.2\sigma$ and $3.1\sigma$ levels for the cases of $s=1$ and $s=3$ respectively.

Finally we consider the general power-law models in which $s$ is treated as a free parameter. We use the dataset of ALL18 to constrain this model, and find that the constraints on $\mu_s$ and $\eta_s$ get diluted compared with the cases of $s=1$ or $s=3$, due to marginalization over $s$, namely, \be\mu_s=-0.334\pm0.186; \ \eta_s=2.090\pm0.904; \ s=2.474 \pm 0.770.\ee In this general case, GR is excluded at around a $2\sigma$ level.

Fig. \ref{fig:fs8} shows the best-fit $f\sigma_8$ of $\Lambda$CDM and power-law models, over-plotted with observational data of RSD. As shown, models with a lower $f\sigma_8$, which means models predicting a weaker gravity, are favored by these recent RSD measurements.

A similar analysis was performed by the Planck collaboration using slightly different power-law functions \citep{MGPLC}, whose conclusion is consistent with ours, \ie, the deviation from GR can reach a $3\sigma$ level (depending on data combinations, see Table 7 in \citealt{MGPLC}). As discussed therein,  besides the RSD measurements, the signal is to some extent due to tensions within $\Lambda$CDM among datasets (see discussions in \citep{ZhaoDE17,PLCWLtension1, PLCWLtension2,VMS16} as well), which may suggest observational systematics, or new physics beyond $\Lambda$CDM. 

\subsubsection{The $z$-binning and PCA}
\label{sec:PCAz}

\begin{figure}[htp]
\includegraphics[scale=0.3]{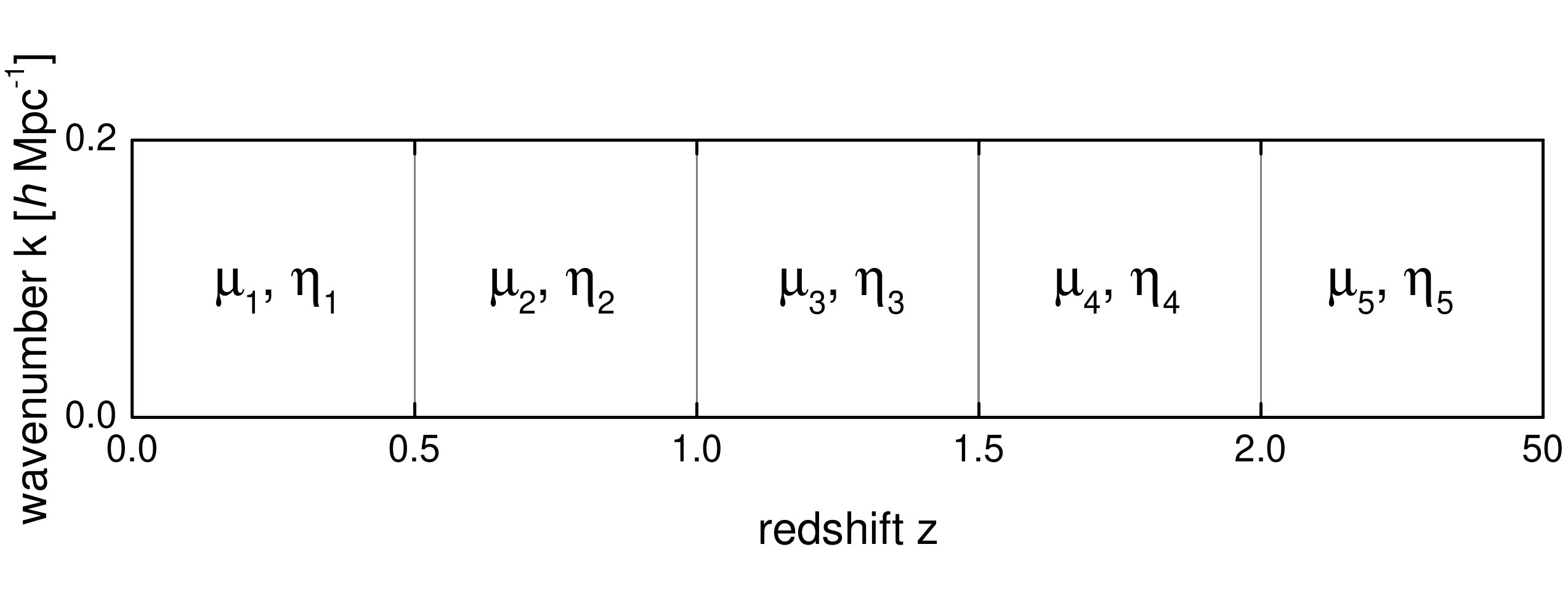}
  \caption{Illustration of the binning scheme in redshift $z$ of the $\mu$ and $\eta$ functions used in this work.}
\label{fig:zbins}	
\end{figure}

In this section, we consider the most general parametrization for scale-independent $\mu$ and $\eta$ functions using piecewise constant bins as free parameters. Given the sensitivity of current observations, we choose the redshift binning as illustrated in Fig. \ref{fig:zbins} \footnote{We assume GR outside the $z$ and $k$ ranges shown in Fig. \ref{fig:zbins}, \ie, $\mu=\eta=1$ if $z>50$ or $k>0.2$ $\hoMpc$.}, thus we have ten MG parameters in total.

\begin{figure*}[htp]
\includegraphics[width=0.93\textwidth]{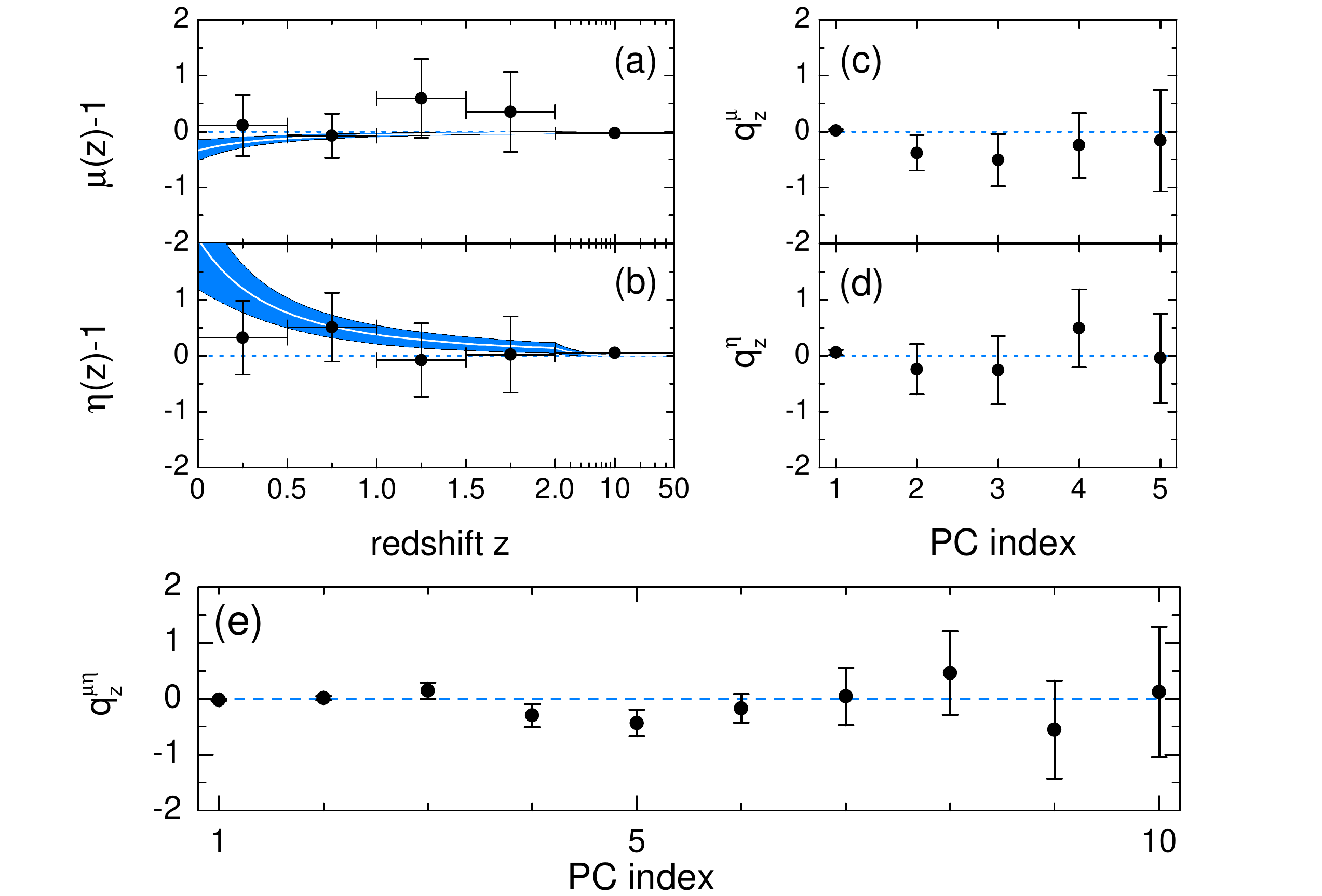}
\caption{Constraints on scale-independent $\mu(z)$ and $\eta(z)$ functions and the associated PCA result derived from ALL18. Panels (a,b): reconstructions of $\mu(z)-1$ (panel a) and $\eta(z)-1$ (panel b) using the redshift bins (data points with 68\% CL error bars). The white curves and blue shaded bands show the mean and 68\% uncertainty of $\mu(z)-1$ and $\eta(z)-1$ reconstructed using the power-law parametrization with $s$ marginalized over. Panels (c,d): mean and 68\% CL uncertainties on coefficients of the principal components (PCs) of the $\mu$ or $\eta$ functions with mutual marginalization (see text for details). Panel (e): mean and 68\% CL uncertainties on coefficients of the PCs of both $\mu$ and $\eta$ functions.}
\label{fig:PCAz}	
\end{figure*}

\begin{table*}[htp]
\begin{centering}
\begin{tabular}{c|c|c|c|c|c}
\hline \hline 
 $\mu_{i}-1$ & $\eta_{i}-1$ & $q_{z}^{\mu}$ & $q_{z}^{\eta}$ & $q_{z}^{\mu\eta}$ (PC1-PC5)& $q_{z}^{\mu\eta}$ (PC6-PC10)\\ \hline 
  $0.110\pm0.546$ & $0.320\pm0.658$ & $0.023\pm0.021$ & $0.059\pm0.045$ & $-0.021\pm0.017$& $-0.175\pm0.256$\\ \hline 
 $-0.074\pm0.396$ & $0.510\pm0.613$ & $-0.381\pm0.314$ & $-0.243\pm0.449$ & $0.013\pm0.038$& $0.041\pm0.515$\\ \hline 
 $0.590\pm0.706$ & $-0.080\pm0.653$ & $-0.507\pm0.468$& $-0.258\pm0.611$ & $0.143\pm0.147$& $0.459\pm0.750$\\ \hline 
 $0.350\pm0.711$ & $0.020\pm0.678$ & $-0.246\pm0.577$ & $0.491\pm0.693$ & $-0.299\pm0.204$& $-0.550\pm0.879$\\ \hline 
  $-0.025\pm0.036$ & $0.050\pm0.048$ & $-0.161\pm0.899$ & $-0.045\pm0.799$ & $-0.436\pm0.237$& $0.120\pm1.170$\\ \hline 
\hline 
 \end{tabular} 
 \caption{Mean and 68\% CL uncertainties on the $\mu_i-1$ and $\eta_i-1$ bins (first two columns on the left) and associated PCA results (the remaining four columns).}
 \label{tab:PCAz} 
 \end{centering}
 \end{table*}

We measure the $\mu$ and $\eta$ bins using the ALL18 dataset, and summarize the results in the left two columns of Table \ref{tab:PCAz} and in panels (a,b) of Fig. \ref{fig:PCAz}. For a comparison with results using other parametrizations, we over-plot a reconstruction of $\mu(z)$ and $\eta(z)$ with 68\% CL uncertainty using the power-law parametrization shown in Eq. (\ref{eq:powerlaw}) with the power index $s$ marginalized over (the blue bands in Fig. \ref{fig:PCAz}), which is in excellent agreement with our binned measurement.

As shown, most of the bins are consistent with the GR prediction except for the $\mu$ bin at $1.0<z<1.5$ (\ie, $\mu_3$ shown in Fig. \ref{fig:zbins}) and for the $\eta$ bin at $0.5<z<1.0$ (\ie, $\eta_2$), both of which exhibit a deviation from GR at approximately $1\sigma$ level. However, as the errors are correlated with each other, it is difficult to interpret the result in a na\"{\i}ve way.

A natural way to interpret the correlated measurements is to perform a Principal Component Analysis (PCA) to decorrelate the covariance matrix of the original parameters, which allows for forming a new set of parameters with a diagonal covariance matrix. The PCA method has been extensively used in cosmology, including implications  in power spectrum measurements \citep{PkPCA1,PkPCA2}, dark energy equation-of-state \citep{wPCA1,wPCA2,wPCA3,wPCA4,wPCA5,wPCA6,ZhaoDE17} and modified gravity parameters \citep{MGPCA1,MGPCA2,MGPCA3,MGPCA4,MGPCA5}.

The essence of the PCA is to diagonalize the covariance matrix $\bf {C_{p}}$ of the original correlated parameters denoted as {\bf p},
\ba {\bf C_{p}} = {\bf W^T \Lambda W}, \ea where {\bf W} is the decomposition matrix and ${\bf \Lambda}$ is the covariance matrix, which is diagonal, for the newly formed uncorrelated parameters {\bf q} = {\bf Wp}. The estimate of q with the associated uncertainty stored in ${\bf \Lambda}$ can identify which modes, \ie, uncorrelated linear combinations of the original parameters, deviate from the expected value given a theory, and how many modes can be constrained by data. 

To investigate the consistency of the $\mu$ or $\eta$ functions with unity, we first perform a PCA on the $\mu$ or $\eta$ bins separately. The PCA result for the $\mu$ bins (with $\eta$ bins marginalized over) and for the $\eta$ bins (with $\mu$ bins marginalized over) are shown in the third and fourth columns and panels (c) and (d) of Fig. \ref{fig:PCAz}. As shown, there are two modes, with Principal Component (PC) indices $2$ and $3$ shown in Fig. \ref{fig:PCAz}, of $\mu$ deviating from the GR value, which is unity, at more than $1\sigma$, while none of the $\eta$ modes show deviation from GR given the uncertainty level. A $\chi^2$ analysis using all the modes shows that the total signal-to-noise ratio (SNR) of $\mu$ and $\eta$ deviating from GR is $2.0\sigma$ and $1.6\sigma$ respectively, based on the improvement in $\chi^2$ only.

To quantify the deviation from GR without distinguishing between  $\mu$ and $\eta$, we perform a PCA on the $\mu$ or $\eta$ bins jointly, and show the result in the last two columns in Table \ref{tab:PCAz} and in panel (e) of Fig. \ref{fig:PCAz}. As illustrated, there are four joint $\mu$ and $\eta$ modes, with PC indices $2,3,4$ and $5$, deviating from GR beyond the uncertainty level, which yields a $3.1\sigma$ signal in total.

The fact that using a large number of bins does not further improve the fitting compared with the power-law case means that the important features in the data can well be resolved by the power-law functions, which is consistent with what we show in panels (a,b) in Fig. \ref{fig:PCAz}. Actually, the PCA result conveys the same message: only $3$ or $4$ modes are needed to reproduce the total variance, which are essentially the degrees of freedom in the power-law functions. 

\subsection{The scale-dependent case}

Now we consider more general cases in which the growth is scale-dependent, \ie, $\mu$ and $\eta$ are functions of both scale and time, namely, 
\ba \mu = \mu(k,a); \ \eta = \eta(k,a). \ea
We then parametrize the $\mu(k,a)$ and $\eta(k,a)$ functions in the framework of the scalar-tensor theories, and use a more general parametrization based on pixelization in both scale and time. 

\subsubsection{A single parameter extension: the $f(R)$ model}
\label{sec:fR}

The $f(R)$ theory \citep{fRreview,HS07,fRPS08,fRdyn} is a special case of the scalar-tensor theory with the following $\mu$ and $\eta$ functions \citep{BZ},\ba\label{eq:BZ}
\mu(a,k) &=& \frac{1+\beta_1 \lambda_1^2 k^2 a^s}{1+\lambda_1^2k^2a^s}, \nn \\
\eta(a,k) &=& \frac{1+\beta_2 \lambda_2^2 k^2 a^s}{1+\lambda_2^2k^2a^s},
\ea where $\beta_1$ and $\beta_2$ (denoting the coupling; dimensionless), $s$ (the power index; dimensionless), $\lambda_1$ and $\lambda_2$ (the length scales; in units of Mpc) are free parameters.

In $f(R)$, \ba \beta_1=4/3; \  \ \ \beta_2=1/2; \ \ \ \lambda_2^2/\lambda_1^2=4/3. \ea
We fix $s=4$ to closely mimic the $\Lambda$CDM model at the background level \citep{AS10}, which leaves only one free parameter, $\lambda_1$, to be constrained. In practice, we vary $\lnB$ together with other cosmological parameters where $B_0 \equiv 2H_0^2 \lambda_1^2/{c^2}$. The Hubble constant $H_0$ and the speed of light $c$ in the equation above make $B_0$ dimensionless, and $B_0=0$ corresponds to the $\Lambda$CDM limit.

\begin{table}[htp]
\begin{centering}
\begin{tabular}{c|c}
\hline\hline
                         &  log$_{10}{B_0}$ (95\% CL upper limit)  \\  \hline
  {PLC+Alam}   &  $-4.276$       \\  \hline
  {PLC+Wang}  &  $-4.913$       \\  \hline
  {ALL17}          &  $-4.950$       \\  \hline
  {ALL18}          &  $-4.932$       \\  \hline\hline
\end{tabular}
\caption{The 95\% CL upper limit on log$_{10}{B_0}$ derived from four data combinations.}
\label{tab:fR}
\end{centering}
\end{table}

\begin{figure}[htp]
\includegraphics[scale=0.7]{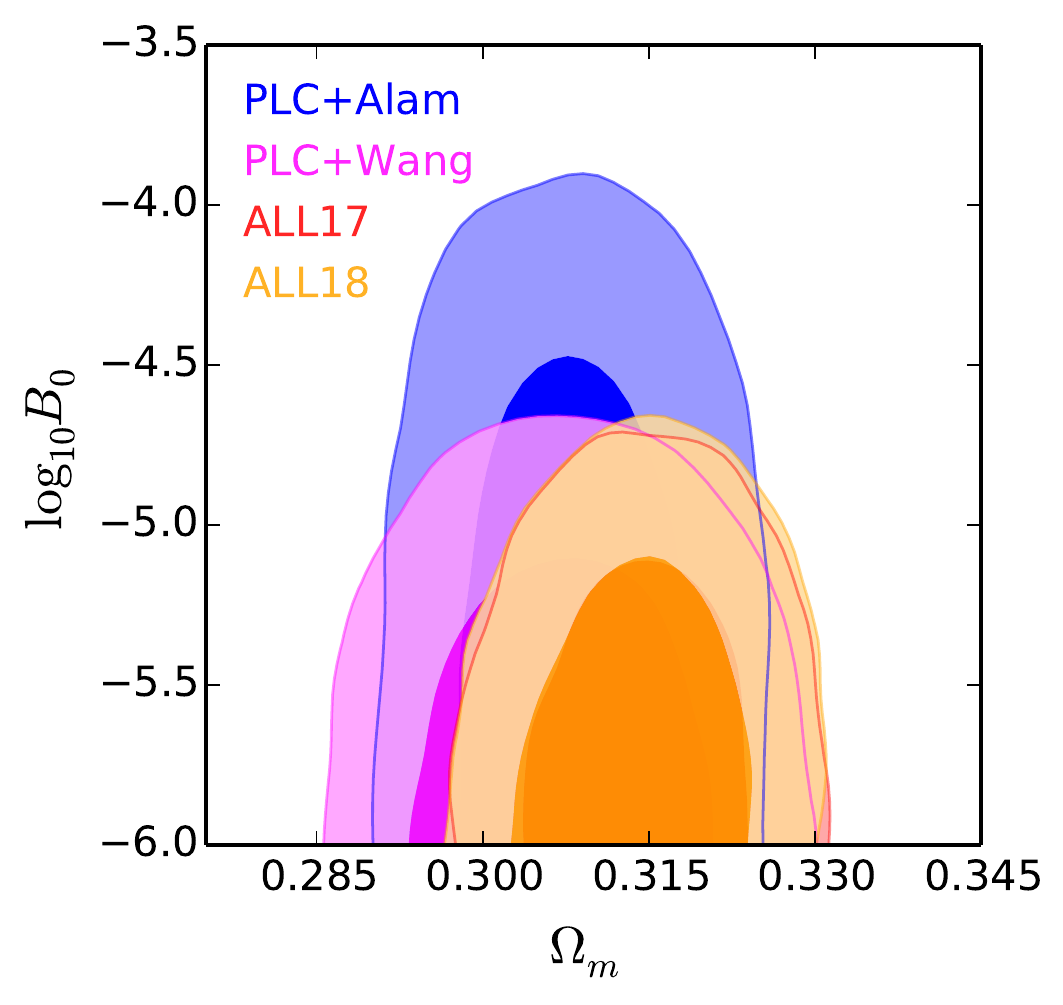}
\caption{68 and 95\% CL contour plots for $\lnB$ and $\Omega_m$ derived from four different data combinations illustrated in the legend.}
\label{fig:fR}
\end{figure}

\begin{table}[htp]
\begin{centering}
\begin{tabular}{c|c|c}
\hline\hline  
Parameter & $s\in[1,4]$ & $s\in[0,10]$  \\ \hline  
$\beta_1$  & $0.974\pm0.033$ & $0.928\pm0.061$  \\ \hline 
$\beta_2$  & $1.349 \pm 0.165$ & $1.647\pm0.296$ \\ \hline\hline
 \end{tabular} 
\caption{The mean and 68\% CL uncertainty on $\beta_1$ and $\beta_2$ derived from ALL18 with two different flat priors on $s$.} 
\label{tab:BZ} 
\end{centering}
\end{table}

\begin{figure} [htp]
\includegraphics[scale=0.3]{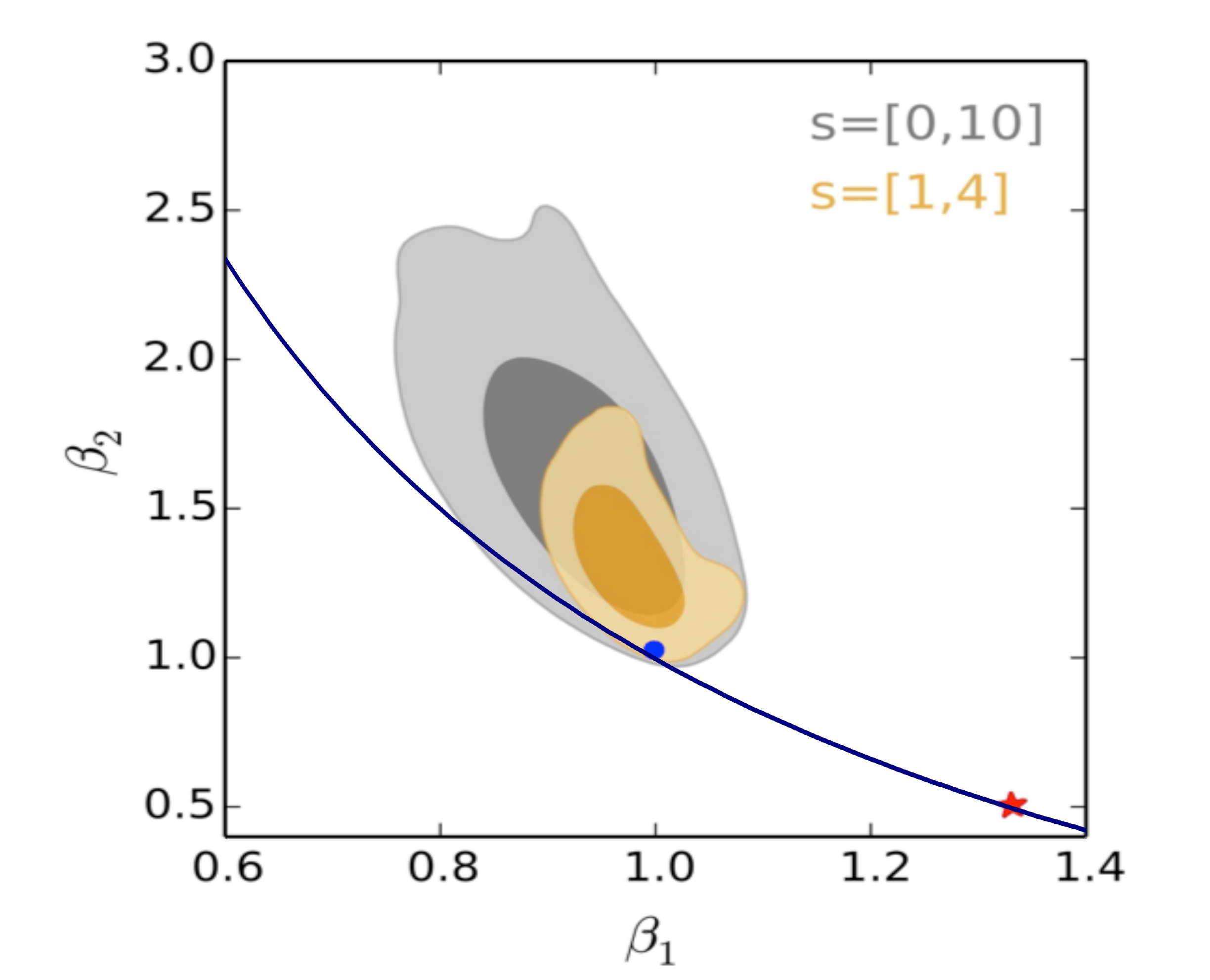}
\caption{ 68 and 95\% CL contour plots for $\beta_1$ and $\beta_2$ derived from ALL18 for two choices of flat priors applied on $s$. The blue solid curve shows the relation of $\beta_2=2/\beta_1-1$. The blue dot and the red star denote the $\Lambda$CDM and the $f(R)$ models respectively.}
\label{fig:BZ}
\end{figure}

\begin{figure}[htp]
\includegraphics[scale=0.35]{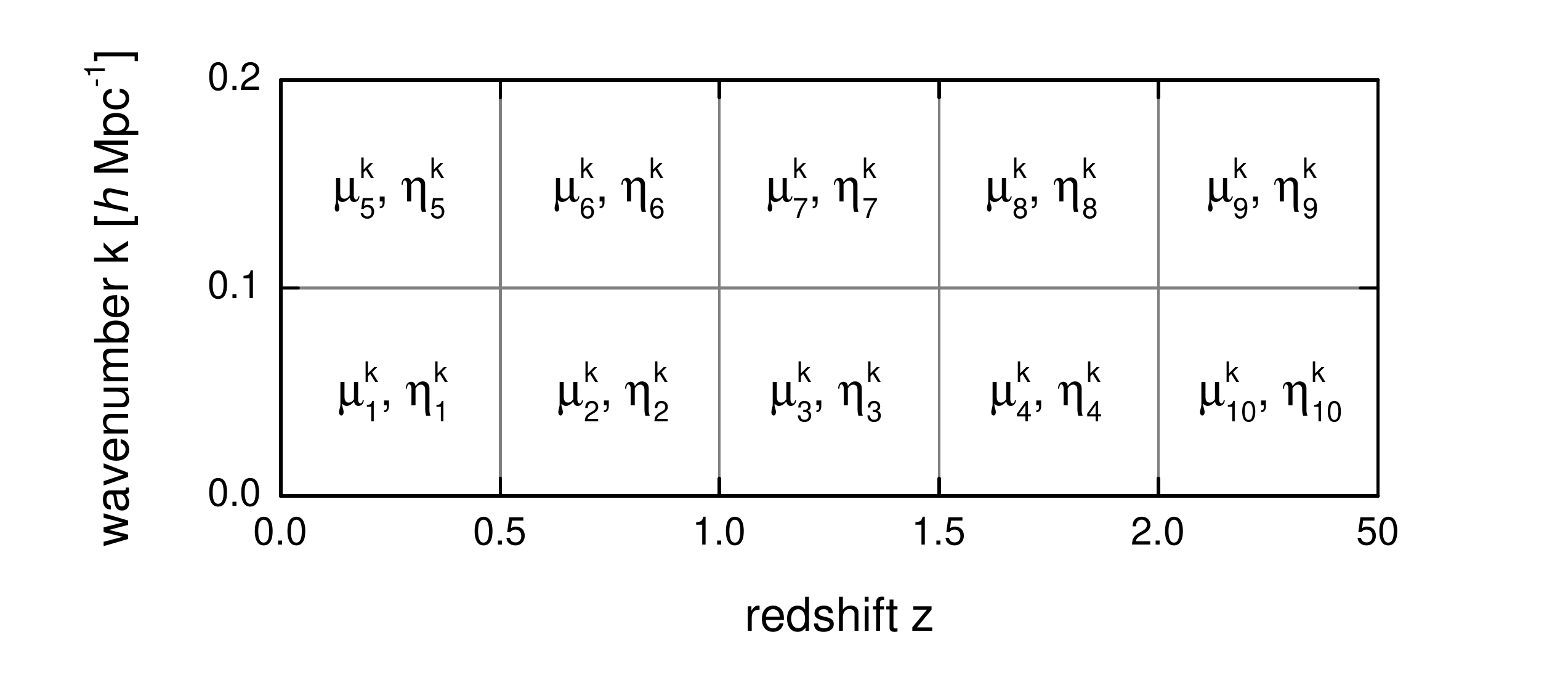}
  \caption{Illustration of the pixelization scheme in $k$ and $z$ of the $\mu$ and $\eta$ functions used in this work.}
\label{fig:grids}	
\end{figure}

The constraint on $f(R)$ gravity using four datasets is shown in Table \ref{tab:fR} and Fig. \ref{fig:fR}. First we notice that the constraint derived from PLC+Wang is much more stringent than that from PLC+Alam, which demonstrates the improvement on MG constraints using tomographic BAO and RSD measurements, as claimed in \cite{Zheng18}. Adding more datasets further improves the constraints, namely, the 95\% CL upper limit of $\lnB$ gets down to $-4.93$ using ALL18, which is tighter than a recent measurement, $\lnB<-4.54$, derived in \citep{EMM18}. This is largely due to the additional information in the tomographic BAO and RSD measurements used for our analysis.

\subsubsection{A five-parameter extension motivated by the scalar-tensor model}

The form of $\mu$ and $\eta$ for general scalar-tensor models is shown in Eq. (\ref{eq:BZ}). Note that for scalar-tensor theories, the following consistency relation holds \citep{MGCAMB1,MGCAMB2}, \ba\label{eq:b1b2} \beta_1 = \frac{\lambda_1^2}{\lambda_2^2}; \ \ \beta_2=\frac{2}{\beta_1}-1. \ea
However these relations are not applied as a constraint in our analysis, but used for a direct comparison with our observational constraint. 

\begin{table*}[htp]
\begin{centering}
\begin{tabular}{c|c|c|c|c|c}
\hline \hline 
 $\mu_{i}^{k}-1$ & $\eta_{i}^{k}-1$ & $q_{kz}^{\mu}$ & $q_{kz}^{\eta}$ & $q_{kz}^{\mu\eta}$ (PC1-PC10)& $q_{kz}^{\mu\eta}$ (PC11-PC20)\\ \hline 
 $0.309\pm0.768$ & $-0.015\pm0.594$ & $-0.024\pm0.023$& $0.036\pm0.042$& $-0.017\pm0.017$ & $-1.212\pm0.694$\\ \hline 
 $-0.177\pm0.341$ & $0.845\pm0.656$ & $-0.060\pm0.199$& $-0.062\pm0.324$& $-0.010\pm0.038$ & $-0.542\pm0.740$\\ \hline 
 $0.696\pm0.597$ & $-0.297\pm0.407$ & $-0.282\pm0.279$& $0.328\pm0.479$& $0.201\pm0.119$ & $0.266\pm0.799$\\ \hline 
 $0.175\pm0.663$ & $0.162\pm0.701$ & $-0.725\pm0.400$& $0.227\pm0.535$& $0.098\pm0.153$ & $-0.317\pm0.901$\\ \hline
 $-0.330\pm0.910$ & $-0.619\pm0.762$ & $0.631\pm0.694$& $-0.779\pm0.675$& $0.121\pm0.172$ & $0.409\pm0.948$\\ \hline 
 $-0.679\pm0.768$ & $-0.700\pm0.880$ & $0.266\pm0.762$& $-0.575\pm0.757$ & $-0.468\pm0.243$ & $-0.090\pm0.970$\\ \hline 
 $-0.314\pm0.971$ & $-0.278\pm1.072$ & $0.003\pm0.830$& $0.042\pm0.767$& $0.054\pm0.250$ & $0.133\pm1.001$ \\ \hline 
 $-0.168\pm1.105$ & $-0.255\pm1.019$ & $0.349\pm0.894$& $0.453\pm0.885$ & $-0.059\pm0.437$ & $0.102\pm1.100$\\ \hline 
 $-0.013\pm0.032$ & $0.034\pm0.043$ & $0.373\pm0.995$& $0.732\pm1.039$ & $-0.384\pm0.476$ & $0.645\pm1.148$\\ \hline 
 $-0.035\pm0.217$ & $-0.001\pm0.604$ & $0.005\pm1.060$& $-0.027\pm1.116$ & $-0.170\pm0.557$ & $0.407\pm1.234$\\
\hline \hline 

 \end{tabular} 
 \caption{The mean and 68\% CL uncertainties on the scale-dependent functions of $\mu(k,z)-1$ and $\eta(k,z)-1$ (first two columns on the left) and associated PCA results (the remaining four columns).}
 \end{centering}
 \label{tab:PCAkz} 
 \end{table*}
 
\begin{table}[htp]
\begin{centering}
 \label{tab:chi2} 
\begin{tabular}{c|c|c|c|c}
\hline\hline  
Model & $\Delta\chi^2$ & SNR & $\Delta N_p$ & SNR$/\Delta N_p$\\ \hline  
$\Lambda$CDM  & $0$ & $0$  & $0$ & $-$\\ \hline 
$\gamma_L$  & $-4.8$ & $2.2$ & $1$ & $2.2$\\ \hline
Power law, $s=1$  & $-12.4$ & $3.5$ & $2$ & $1.7$\\ \hline
Power law, $s=3$  & $-12.8$ & $3.6$ & $2$ & $1.8$\\ \hline
Power law, $s$ floating  & $-12.4$ & $3.5$ & $3$ & $1.2$\\ \hline
BZ model, $s\in[0,10]$  & $-11.2$ & $3.3$ & $5$ & $0.66$\\ \hline
BZ model, $s\in[1,4]$  & $-12.2$ & $3.5$ & $5$ & $0.7$\\ \hline
$z$-binning & $-10.6$ & $3.3$ & $10$ & $0.33$\\ \hline
$k,z$-pixelization & $-17.0$ & $4.1$ & $20$ & $0.21$\\ \hline
\hline
 \end{tabular} 

 \caption{The improved $\chi^2$ ($\Delta\chi^2$), the signal-to-noise ratio calculated using the improved $\chi^2$ alone (${\rm SNR}\equiv\sqrt{|\Delta\chi^2|}$), the additional parameters to the $\Lambda$CDM model ($\Delta N_p$), and the signal-to-noise ratio per additional parameter (SNR$/\Delta N_p$) for the constraint on MG models studied in this work using the ALL18 dataset.}
 \end{centering}
 \end{table}
 
 \begin{figure*}[htp]
\includegraphics[width=0.93\textwidth]{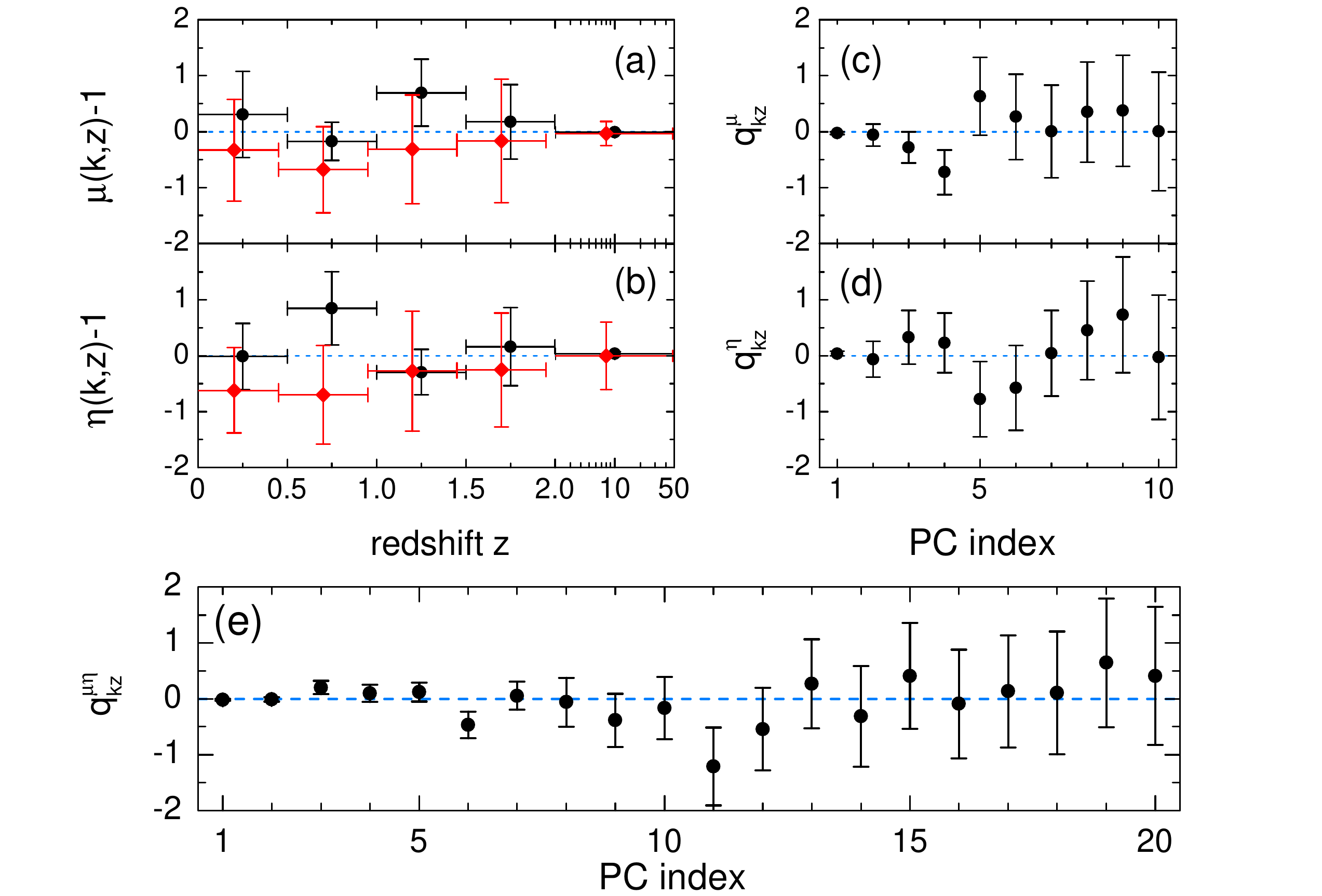}
  \caption{Constraints on scale-dependent $\mu(k,z)$ and $\eta(k,z)$ functions and the associated PCA result derived from ALL18. Panels (a,b): reconstructions of $\mu(k,z)-1$ (panel a) and $\eta(k,z)-1$ (panel b) using the $z$ and $k$ pixels (data points with 68\% CL error bars). The black circles with error bars and red diamonds with error bars represent pixels within $k\in[0,0.1]$ and $k\in[0.1,0.2]$ respectively. Panels (c,d): the mean and 68\% CL uncertainties on coefficients of the PCs of the $\mu$ or $\eta$ functions with mutual marginalization. Panel (e): the mean and 68\% CL uncertainties on coefficients of the principal components (PCs) of both $\mu$ and $\eta$ functions.}
\label{fig:PCAkz}	
\end{figure*}

It is worth noting that a large $s$ can make other parameters trivial in the joint parameter estimation, thus a prior on $s$ is needed. In this work, we make two choices of the flat prior for $s$. One is motivated by scalar-tensor theories, which is $s\in[1,4]$ \citep{MGCAMB2,AS10}, with another one being more conservative, namely, $s\in[0,10]$.

We show the constraints of $\beta_1$ and $\beta_2$ derived from ALL18 with all other parameters marginalized over in Fig. \ref{fig:BZ} and Table \ref{tab:BZ}. As shown in both cases, GR ($\beta_1=\beta_2=1$) is consistent with data at 68\% CL, and the scalar-tensor theory prediction, Eq (\ref{eq:b1b2}), is allowed within the 68\% CL uncertainty. However, the $f(R)$ model discussed in Section \ref{sec:fR} with $\beta_1=4/3, \ \beta_2=1/2$ is strongly disfavored by data. This is understandable as we have seen from the power-law case in Sec. \ref{sec:powerlaw} (see Fig. \ref{fig:fs8}) that data favor a weaker gravity, while in $f(R)$, gravity is always stronger than that in GR.

\subsubsection{The $k,z$-pixelization and PCA}

We parametrize the functions of $\mu$ and $\eta$ using pixels in the $(k,z)$ plane as illustrated in Fig. \ref{fig:grids}, constrain the pixels using the ALL18 dataset, and present the result in Table \ref{tab:PCAkz} and Fig \ref{fig:PCAkz} in a similar way as we did for the $k$-independent case in Sec. \ref{sec:PCAz}.

Looking at the constraints on the pixels shown in the left two columns in Table \ref{tab:PCAkz} and in panels (a,b) in Fig. \ref{fig:PCAkz}, we find that pixels $\mu_3^k,\eta_2^k$ and $\eta_6^k$, as denoted in Fig. \ref{fig:grids}, show a deviation from GR at more than $1\sigma$ level, and interestingly, the $\eta$ function at $z\in[0.5,1.0]$ shows a signal of scale-dependence at around $2\sigma$ level. 

A PCA on $\mu$ and $\eta$ pixels with the other parameters marginalized over shows that there are three (two) $\mu$ ($\eta$) modes deviating from GR beyond the uncertainty, which corresponds to a $2.6\sigma$ and $2.0\sigma$ signal respectively. A PCA on all the $\mu$ and $\eta$ pixels jointly reveals four modes, with PC indices $3,5,6, and\; 11$, deviating from GR noticeably, making a total signal at the level of $3.7\sigma$. This means that only a small number of degrees of freedom is required to capture the feature in the data, which is consistent with the scale-independent case.

\section{Conclusion and Discussion}  
\label{sec:conclusion}

Theoretical and observational approaches have been developing in order to test the validity of postulating GR on cosmological scales, which is a significant extrapolation of our knowledge of gravity from scales within the solar system. Observational tests of theoretical models thus play a crucial role in search for the ultimate theory of gravity governing the observed cosmic acceleration. As a large number of modified gravity theories have been proposed (see reviews of \citealt{MGreview1,MGreview2,MGreview3,MGreview4}), it is efficient to perform observational gravity tests following a phenomenological approach. 

In this work, we parametrize the effect of modified gravity using two functions $\mu$ and $\eta$ on linear scales, which are generically dependent on both time and scale, describing the effective Newton's constant and the gravitational slip respectively, and use the latest observational data to constrain parameters for these two functions. 

By assuming that $\mu$ and $\eta$ only depend on time to start with, we further parametrize them using the gravitational growth index $\gamma_L$, power-law functions and piece-wise constant bins progressively. We find no signal of modified gravity from current observations using $\gamma_L$, which is a one-parameter extension of $\Lambda$CDM, but see a significant deviation from GR (at around $3\sigma$ level) using the power-law parametrization (a two-parameter extension). Using a more general parametrization with piecewise constants in redshifts (a ten-parameter extension), we find that the significance stays at the same level, signaling that the important features in the data, which can be described by a scale-independent growth, can well be extracted using power-law functions for $\mu$ and $\eta$.

We then further explore more general cases in which both $\mu$ and $\eta$ depend on time and scale. We parametrize these two functions in frameworks of $f(R)$ gravity (a one-parameter extension of GR), scalar-tensor theory (a five-parameter extension), and using pixels (a twenty-parameter extension). We find no significant deviation from GR in $f(R)$ or in the scalar-tensor models, but a deviation at a $3.7\sigma$ level is revealed when using pixels. We caution that the signal-to-noise ratio quoted here is computed using the improved $\chi^2$ of the fitting, thus is not sufficient for a model selection. In Table \ref{tab:chi2}, we show the improvement in the $\chi^2$, as well as that normalized by the number of additional parameters, for the MG models. As shown, the most `parameter-economic' model, in which SNR/$\Delta N_p$ get maximized, is the $\gamma_L$ model, which shows no deviation from GR. The power-law models with $s=1$ and $s=3$ are slightly less parameter-economic, but a significant deviation from GR is seen in such models. An evaluation of the Bayesian Evidence is needed for a formal model-selection, which is left for a future work.

The signal we find in this work is to some extent due to tensions among datasets on cosmological scales within the $\Lambda$CDM model, which has been investigated by the community. This could be due to contaminations from unknown systematics in the observations, or a sign of new physics, which can be further studied by complementary GR tests on non-linear scales \citep{AstroMG0,AstroMG1,AstroMG2,AstroMG3,AstroMG4,profileMG,peakMG,voidMG,EG1,EG2,FLZ17}. Forthcoming large astronomical surveys, including Dark Energy Spectroscopic Instrument (DESI) \citep{DESI}, Prime Focus Spectrograph (PFS) \citep{PFS} and the Euclid satellite \citep{Euclid}, will provide rich observational data for GR tests across a large range of scales.

\acknowledgments

We thank Yuting Wang, Xiao-Dong Li, Eva-Maria Mueller and Will Percival for discussions. This work is supported by the National Key Basic Research and Development Program of China (No. 2018YFA0404503), and by NSFC Grants 11720101004, 11673025 and 11711530207. This research used resources of the SCIAMA cluster supported by University of Portsmouth.


 
\bibliography{draft}



\end{document}